\documentclass[aps,preprint,amsmath,amssymb]{revtex4}

\usepackage{graphicx}
\usepackage{amsmath}
\usepackage{bm}
\usepackage{multirow}
\newcommand{\be}{\begin{equation}}
\newcommand{\ee}{\end{equation}}
\newcommand{\bear}{\begin{eqnarray}}
\newcommand{\eear}{\end{eqnarray}}

\begin{document}

\title{Why there is no spin-orbit inversion in heavy-light mesons?}

    \author{Ian Woo Lee and Taekoon Lee}
\email[]{tlee@kunsan.ac.kr}
\affiliation{Department of Physics, Kunsan National University,
Kunsan 573-701, Korea}

\begin{abstract}
We show that the absence of  spin-orbit inversions in  heavy-light
mesons can be explained by the chiral radiative 
corrections in the  potential model. A new potential model 
estimate is given of the masses for P-wave bottom mesons. 
\end{abstract}


\maketitle
\section{Introduction}
It was suggested long ago by Schnitzer that the strong spin-orbit
interaction of the scalar confining potential would lead to spin-orbit
inversions in P-wave heavy-light mesons, with the claim
that their observation would confirm the scalar nature of the 
confining potential \cite{schnitzer}. This spin-orbit inversion
was later reaffirmed in studies with  more sophisticated  potential
models \cite{inversion1,inversion2}.
However, contrary to these studies,  the observed
masses of the P-wave charmed mesons
do not exhibit  spin-orbit inversion.

In this paper we show that if the one loop chiral  corrections
 are taken into account  then  spin-orbit
inversions disappear, and the experimental data can be understood  
within the potential model. Our result suggests that 
the absence of  spin-orbit inversions
in the observed P-wave mesons
should not be interpreted as the failure
of  potential
model or the confining potential 
be  of non-scalar type, but, rather, should be regarded 
as a support for the
potential model that is  augmented by 
radiative corrections.

\section{The model}
We use the  relativistic potential model of heavy-light system 
\cite{roberts,pe} based on the chiral quark model, with the axial
coupling of the light mesons put in explicitly. The Lagrangian reads 
\bear
{\cal L} &=& \Psi^\dagger (i\partial_0 -H) \Psi +
g_A \bar\Psi \not\!\! A \gamma_5 \Psi +{\cal L}_{\Pi}\nonumber\\
              &\approx& \Psi^\dagger (i\partial_0 -H) \Psi+\frac{
g_A}{2f_{\pi}}\bar\Psi_i\gamma^\mu\gamma_5\Psi_j
	      \partial_\mu\Pi_{ij} +{\cal L}_{\Pi}\,,
\label{lagran}
\eear
where $\Pi=\sum_{a=1}^{8}\pi^a \lambda^a$ and 
$\Psi_i=(u,d,s)$ are the light octet mesons and the light
quark fields, respectively,  and  ${\cal L}_{\Pi}$ denotes the 
chiral Lagrangian
for the  light mesons, and
\bear
A_\mu=\frac{i}{2}(\xi^\dagger\partial_\mu\xi-
\xi\partial_\mu\xi^\dagger)
\eear
with $\xi=e^{i\Pi/2f_\pi}$. 
The Hamiltonian $\rm H$ is 
given by
\bear
\rm H=\rm H_0+\frac{1}{\rm M} \rm H_1 + \cdots 
\eear
where $\rm M$ denotes the heavy quark mass.
The leading Hamiltonian $\rm H_0$ in the heavy quark 
mass expansion reads
\bear
\rm H_0= \gamma^0(-i \not\!\nabla + {\bf m}) + V(r)
\eear 
where ${\bf m}={\rm m}_i\delta_{ij}$ and  $V(r)$ 
denote the constituent quark masses and the potential, respectively.

The spectrum  of the resonances in conventional potential model
is obtained by solving the Dirac
equation from $\rm H_0$, followed by time-independent
perturbations of the 
subleading terms. The free parameters of the model are fixed by
fitting the predicted masses to those of the observed 
resonances.  The masses obtained this way do not agree well with
the P-wave charmed mesons, for instance the mass of 
$D_s(2317)$ is much lower than the potential model prediction.

Since the heavy-light mesons
are chirally active  the  masses in the potential model
get chiral 
radiative corrections via the axial coupling. We pointed out in
\cite{lee} that the one loop corrections are sizable,
comparable to the $1/{\rm M}$ corrections, and must be
incorporated in potential model calculations. 

The chiral radiative corrections were successful in understanding
why  $D_s(2317)$  and  $D(2308)$ have such close masses.
In general, potential models predict, roughly,
about 100 MeV larger masses
for strange states over their non-strange counterparts, and 
 the gap defined as
\bear
{\rm gap}\equiv 
[m(D(0^+))-m(D(0^-))]-[m(D_s(0^+))-m(D_s(0^-))]
\eear
almost vanishes in potential model, whereas, experimentally, the
gap is about $95$ MeV. When the chiral radiative corrections are taken
into account, however, the potential model predicts a gap that
is not only consistent with the experimental value but also
insensitive to the ultraviolet (UV) 
cutoff, with the main contribution to the gap
coming from the
low energy region of about $250$ MeV, far down the cutoff \cite{lee}.

The potential model being nonrenormalizable
the loop corrections depend on the regularization scheme chosen.
In Ref. ~\cite{lee} we introduced a three-momentum UV
cutoff regularization, and here we employ the same scheme. 
Our potential 
model with radiative corrections is thus the relativistic potential model
 (\ref{lagran}) with the three-momentum cutoff regularization.

\section{Spin-orbit inversion}

In this paper we take the model in 
Ref. ~\cite{pe} as a reference potential model of
conventional type.
The model predicts spin-orbit inversions in $D, D_s$ as well
as in $B, B_s$ mesons.  
For example, with the meson states denoted by $H(l,j,J)$, the 
P-wave states
$D_s(1,\frac{1}{2},1)$ and $D_s(1,\frac32,1)$,
which have the same hyperfine splitting,
are predicted to have masses 2605 MeV and 2535 MeV, respectively.
However, the experimental values   are 
2460 MeV and 2535 MeV, respectively,
and there is no spin-orbit inversion.
Spin-orbit inversions are also predicted  in D-wave states  as well.
For instance,
the masses of $D_s(2,\frac32,2)$ and $D_s(2,\frac52,2)$ are,
respectively, 2953 MeV and  2900 MeV. The spin-orbit inversions
in D-wave states have not been tested yet, but as we shall see, 
when the radiative corrections are
taken into account
there should be no spin-orbit inversions in D-wave states.

To understand the absence of spin-orbit inversions
 within the potential model one must consider a new
effect that origins from the terms in the Hamiltonian that
do not depend on the heavy quark mass, since 
the spin-orbit inversions survive in the heavy-quark limit.
We shall show
that the  chiral radiative corrections can be such a 
new effect. 

\section{Radiative corrections for P and D-wave states}
 In Ref. \cite{lee} we calculated the loop 
 corrections for S and P-wave states
 with $j=\frac{1}{2}$ to estimate the mass gap. In this paper we
 extend the calculation to P-wave 
 states with $j=3/2$ as well as D-wave states
 with $j=3/2$ and $ 5/2$.
 The loop corrections can be obtained similarly as in Ref. \cite{lee},
 to which we refer the readers for details.
 The loop amplitudes can be most conveniently organized by decomposing
 the plane wave of the internal light-meson
 propagator into spherical harmonics.
 A loop amplitude is then given as a sum over the angular 
 momentum quantum
 numbers $l_\pi,m_\pi$ and $j_{\bf n},m_{\bf n}$ of the light
 meson and internal state, respectively, in the Feynman diagram.
 The summation over $m_\pi,
 m_{\bf n}$ can be performed exactly and  the 
 final result for the one loop correction to the energy of
 the  P and D-wave 
 states, labeled by the quantum numbers ${\bf m}=(n,l,j,m_j,q)$,
 can be written as
\bear
   \Delta E_{\bf m}^{\rm loop} = \sum_{\bf n}\sum_{\pi,l_\pi} \zeta_\pi 
                      J({\bf m},{\bf n},l_\pi) 
		      C(l_\pi,l_{\bf n},j_{\bf n}),  
 \label{loopcorrection}
 \eear
where the factor $C(l_\pi,l_{\bf n},j_{\bf n})$ is given by
\bear
 C(l_\pi,l_{\bf n},j_{\bf n})= j_{\bf n}+\frac{1}{2}
  \label{delE1/2P}
\eear
for $l_{\bf m}=1,j_{\bf m}=1/2$, and 
\bear
 C(l_\pi,l_{\bf n},j_{\bf
 n})&=&\left(\frac{3l_\pi(l_\pi-1)}{2(2l_\pi-1)}\delta_{l_{\bf n},l_\pi-2}
 +\frac{l_\pi(l_\pi+1)}{2(2l_\pi+3)}\delta_{l_{\bf
 n},l_\pi}\right)\delta_{j_{\bf n},l_{\bf n}+\frac{1}{2}} +\nonumber \\
 &&\left(\frac{l_\pi(l_\pi+1)}{2(2l_\pi-1)}\delta_{l_{\bf n},l_\pi}
 +\frac{3(l_\pi+1)(l_\pi+2)}{2(2l_\pi+3)}\delta_{l_{\bf
 n},l_\pi+2}\right)\delta_{j_{\bf n},l_{\bf n}-\frac{1}{2}}
 \label{delE3/2P}\eear
 for $l_{\bf m}=1,j_{\bf m}=3/2$, and
\bear
 C(l_\pi,l_{\bf n},j_{\bf
 n})&=&\left(\frac{l_\pi(l_\pi+1)}{2(2l_\pi-1)}\delta_{l_{\bf n},l_\pi-1}
 +\frac{3(l_\pi+1)(l_\pi+2)}{2(2l_\pi+3)}\delta_{l_{\bf
 n},l_\pi+1}\right)\delta_{j_{\bf n},l_{\bf n}+\frac{1}{2}} +\nonumber \\
 &&\left(\frac{l_\pi(l_\pi+1)}{2(2l_\pi+3)}\delta_{l_{\bf n},l_\pi+1}
 +\frac{3l_\pi(l_\pi-1)}{2(2l_\pi-1)}\delta_{l_{\bf
 n},l_\pi-1}\right)\delta_{j_{\bf n},l_{\bf n}-\frac{1}{2}}
\label{delE3/2D} \eear
  for $l_{\bf m}=2,j_{\bf m}=3/2$,
and
\bear
 C(l_\pi,l_{\bf n},j_{\bf
 n})&=&\left(\frac{l_\pi(l_\pi-1)(l_\pi+1)}{(2l_\pi-1)(2l_\pi+3)}
 \delta_{l_{\bf n},l_\pi-1}
 +\frac{l_\pi(l_\pi+1)(l_\pi+2)}{2(2l_\pi+3)(2l_\pi+5)}
 \delta_{l_{\bf n},l_\pi+1}  \right. \nonumber\\
 && \left.\quad + \frac{5l_\pi(l_\pi-1)(l_\pi-2)}{2(2l_\pi-1)(2l_\pi-3)}
 \delta_{l_{\bf n},l_\pi-3}\right)\delta_{j_{\bf n},l_{\bf n}+\frac{1}{2}}
 +\nonumber \\
 &&\left(\frac{l_\pi(l_\pi-1)(l_\pi+1)}{2(2l_\pi-1)(2l_\pi-3)}
 \delta_{l_{\bf n},l_\pi-1}
 +\frac{l_\pi(l_\pi+1)(l_\pi+2)}{(2l_\pi-1)(2l_\pi+3)}
 \delta_{l_{\bf n},l_\pi+1} \right. \nonumber \\
 && \left. \quad +\frac{5(l_\pi+1)(l_\pi+2)(l_\pi+3)}{2(2l_\pi+3)(2l_\pi+5)}
 \delta_{l_{\bf n},l_\pi+3}\right)\delta_{j_{\bf n},l_{\bf n}-\frac{1}{2}}
\label{delE5/2D} \eear
 for $l_{\bf m}=2,j_{\bf m}=5/2$. 
Here, following the notation in ref. \cite{lee},  
\begin{eqnarray}
 J({\bf m},{\bf n},l_\pi)
    &=& -\frac{g_A^2}{8f_\pi^2}  
      \int_0^\infty \frac{k^2 dk}{(2\pi)^3 E_\pi} 
   \left[ 
      (E^0_{\bf n} - E^0_{\bf m})|\rho^{(1)}_{\bf mn}(|\vec{k}|,l_\pi)|^2
   \right.\nonumber\\
 &&\left.
     \hspace{-2.3cm}+ 2{\rm Re}[\rho^{(1)}_{\bf mn}(|\vec{k}|,l_\pi)
            \rho^{(2)*}_{\bf mn}(|\vec{k}|,l_\pi)]
     + \frac{|\rho^{(2)}_{\bf mn}(|\vec{k}|,l_\pi)|^2}
            {E_\pi - E^0_{\bf m} + E^0_{\bf n}-i\epsilon}
   \right]
  \label{J_mn_lm}
\end{eqnarray}
with
\begin{eqnarray}
   \rho^{(1)}_{{\bf m},{\bf n}}(|\vec{k}|,l_\pi) 
  &=&      \sqrt{4\pi} \int^\infty_0 r^2 dr 
       (f^{}_{{\bf m}}(r) g^{}_{{\bf n}}(r) - 
       f^{}_{{\bf n}}(r) g^{}_{{\bf m}}(r)) j_{l_\pi}^{}(kr),\\
   \rho^{(2)}_{{\bf m},{\bf m}}(|\vec{k}|,l_\pi) 
   &=& 
      \sqrt{4\pi} \int^\infty_0 r^2 dr 
       (f^{}_{{\bf m}}(r) g^{}_{{\bf n}}(r) 
       +	 f^{}_{{\bf n}}(r) g^{}_{{\bf m}}(r)) \nonumber \\
&& \hskip 7em \times 
({\rm m}_{\bf m} + {\rm m}_{\bf n} + 2V_s)j_{l_\pi}^{}(kr)\,,
\end{eqnarray}
where $E_\pi=\sqrt{k^2+{\rm m}_\pi^2}$, with ${\rm m}_\pi$ 
denoting the light-meson masses, and
$E^0_{\bf m,n}$ and $f_{\bf m,n}(r), g_{\bf m,n}(r)$ are the eigenvalues and
radial wave functions of the
eigenstates of $H_0$, respectively,
$j_{l}(kr)$ denotes the spherical Bessel functions, and  in Eq.
(\ref{loopcorrection})  $\zeta_{\pi}$ denotes
the ${\rm SU}(3)_{\rm flavor}$ factors.

The selection rule for the internal states, labeled by {\bf n},
and the quantum number $l_\pi$ 
in the summation
in Eq. (\ref{loopcorrection}) is given by 
\be
l_\pi+l_{\bf m}+l_{\bf n}={\rm odd \,\,\, integer}
\ee
and
\be
 |j_{\rm m} -l_\pi|\leq j_{\bf n} \leq  j_{\rm m} +l_\pi\,,
\ee
which come from parity and angular momentum conservations at the
vertex in the loop diagram. As an example, the possible internal states
in corrections for a D-wave state 
with $j=5/2$ are listed in Table \ref{table1} for low values of
$l_\pi$. 
\begin{table}[t]
\begin{tabular}{cccccccc}
$l_\pi$=0 && $^\frac52$F&&&&&\\
$l_\pi$=1 && $^\frac32$D & $^\frac52$D & $^\frac72$G &&&\\
$l_\pi$=2 && $^\frac12$P & $^\frac32$P & $^\frac52$F & $^\frac72$F & $^\frac92$H &\\
$l_\pi$=3 && $^\frac12$S & $^\frac32$D & $^\frac52$D & $^\frac72$G & $^\frac92$G & $^\frac{11}{2}$I\\
$l_\pi$=4 && $^\frac32$P & $^\frac52$F & $^\frac72$F & $^\frac92$H & $^\frac{11}{2}$H & $^\frac{13}{2}$J
\end{tabular}
\caption{Internal states $^jL$ allowed at 
given $l_\pi$ in  corrections for a state with
 $j=5/2, l=2$.}
\label{table1}
\end{table}

\section{Modified energy levels}

We can now see how the spin-orbit inversions are affected
by the radiative corrections
given in the previous section. For this we focus on the 
effects of the loop
corrections on the existing potential model
predictions in Ref. \cite{pe}.

The new energy levels, denoted by $\bar {E}_{\bf m}$, that include 
the radiative
corrections can be related to the energy levels of the conventional
potential model by
\bear
\bar{E}_{\bf m}=E_{\bf m} +\delta E^0_{\bf m} +
\delta E^{\rm loop}_{\bf m}
\label{relation}
\eear
where $E_{\bf m}$ denotes the conventional energy levels which contain
the leading order level $E^0_{\bf m}$ as well 
as the $1/{\rm M}$ corrections, and
$\delta E^0_{\bf m}$ denotes the shift in the
leading order level  $E^0_{\bf m}$
caused by the shift in the fitted values of the parameters of 
the model, which was induced by the introduction of
radiative corrections $\delta E^{\rm loop}_{\bf m}$ in 
the fitting of the
parameters.
The relation  (\ref{relation})
is valid to the
leading order of the loop corrections.

Now applying the relation (\ref{relation}) to states with differing  $j$ but
otherwise same quantum numbers in the heavy-quark limit we have
\bear
\bar{E}_{j}-\bar{E}_{j'}=E_j-E_{j'}
+\delta E^{\rm loop}_{j}-\delta E^{\rm loop}_{j'}
+\delta (E^0_j-E^0_{j'})\,,
\label{diffrel}
\eear
where the quantum numbers other than $j$ are suppressed.
Obviously, $\delta E^0_j$ cannot be obtained without 
refitting the parameters of the model, which is beyond the scope of the
present paper, but, fortunately, the differences
$\delta (E^0_j-E^0_{j'})$ are expected to be
small since $E^0_j-E^0_{j'}$ are already small, between 20 and 50 MeVs,
for those states
considered here. We can thus ignore the last term in (\ref{diffrel})
to obtain
\bear
\bar{E}_{j}-\bar{E}_{j'}=E_j-E_{j'}
+\delta E^{\rm loop}_{j}-\delta E^{\rm loop}_{j'}\,.
\label{diffrel2}
\eear
This equation with the loop corrections is our main tool for the investigation
of the spin-orbit inversion.

Let us now focus on the
spin-orbit inversions in P-wave 
states. Using the result in the previous
section we obtain numerical values for the loop 
corrections $\delta E^{\rm loop}_{l,j,q}$
for strange and non-strange mesons in
P and D-wave states (see Table \ref{tab.loop}).
The numbers were obtained using the fitted  parameter values given in
Ref. \cite{pe} and  the UV cutoff put  at 700 MeV.
In our model the UV cutoff is a  parameter that should be fixed along with 
other potential model parameters  by a fitting similar to that
performed in Ref. \cite{pe}.
In the absence of the fitting we here  pick up the preferred
value for the
cutoff suggested in Ref. \cite{lee} where the value 700 MeV was found to
give a reasonable size for the loop corrections.

Looking on Table \ref{tab.loop}
we notice that the magnitudes of the loop corrections are 
larger for states with smaller $j$, and this feature
will be crucial for understanding 
the absence of spin-orbit inversions.

\begin{table}[t]
\begin{tabular}{cccccccc}\hline\hline
$\delta E^{\rm loop}_{1,\frac{1}{2},d}$&$\delta E^{\rm loop}_{1,\frac{3}{2},d}$&$\delta
E^{\rm loop}_{1,\frac{1}{2},s}$&$\delta E^{\rm loop}_{1,\frac{3}{2},s}$ 
&$\delta E^{\rm loop}_{2,\frac{3}{2},d}$&$\delta E^{\rm loop}_{2,\frac{5}{2},d}$&$\delta
E^{\rm loop}_{2,\frac{3}{2},s}$&$\delta E^{\rm loop}_{2,\frac{5}{2},s}$\\ \hline
-261&-183&-344&-181&-257&-184&-275&-184 \\
\hline\hline
\end{tabular}
\caption{Loop corrections $\delta E^{\rm loop}_{l,j,q}$ for P
and D-wave states
in the lowest radial excitations. (Units are in MeV.)}
\label{tab.loop}
\end{table}

We shall first consider the effects of the loop corrections
on the P-wave   $D_s$ mesons. In the following all the states,
labeled as before by $H(l,j,J)$, are in
their lowest radial excitations. We
shall  assume that the modified energy level for the state
$D_s(1,\frac{1}{2},0)$ coincides with the experimental mass
of $D_s(2317)$, and then estimate
the masses of $j=1/2$ and $3/2$ states. Reading the values
of the conventional energy levels from Ref. \cite{pe} and loop
corrections from Table ~\ref{tab.loop} we find 
the following new energy levels of the states 
related by Eq. (\ref{diffrel2}):
\bear
\bar E_{D_s(1,\frac{1}{2},1)}&=&2435 \,\,\,{\rm MeV} \,,\nonumber \\
\bar E_{D_s(1,\frac{3}{2},1)}&=& 2527 \,\,\,{\rm MeV} \,,\nonumber \\
\bar E_{D_s(1,\frac{3}{2},2)}&=& 2573 \,\,\,{\rm MeV}\,.
\eear
Comparing this result with the experimental values 2460 MeV,
2535 MeV, and 2573 MeV, respectively, we see there is good 
agreement between the
new levels and data, and there are no longer  spin-orbit inversions. 

Using the same procedure  we can obtain the modified energy
levels for P-wave  $D$ mesons as well, and the result is summarized in
Table \ref{c-mesons}. We assumed the mass of
$D(1,\frac{1}{2},0)$ coincides with the Belle measurement $2308\pm36$
for $D(0^+)$ \cite{belle}. The Belle mass for
$D(1,\frac{1}{2},1)$ is $2427\pm 42$ \cite{belle}, and considering
the large  uncertainty in the measured value
we find our estimate
is consistent with data in $D$ mesons as well. We note, however, that
if we use the FOCUS
value $2407\pm41$ for $D(0^+)$  mass  \cite{focus}  then
our estimate is no longer consistent with data.

We can  now use Eq. (\ref{relation}), along with Eq. (\ref{diffrel2}), 
to estimate the masses of the P-wave bottom mesons.
We shall first compute the mass for $B_s(1,\frac{1}{2},0)$, which is
the  counterpart of $D_s(2317)$. 
Since the loop corrections are independent of the heavy quark mass we see that
the last two terms in Eq. (\ref{relation}) to be heavy-quark mass
independent, so
we get
\be
\bar{E}_{B_s(l,j,J)}-E_{B_s(l,j,J)}=\bar{E}_{D_s(l,j,J)}-E_{D_s(l,j,J)}\,.
\label{charmbottomrel}
\ee
Identifying again $\bar{E}_{D_s(1,\frac{1}{2},0)}$ with the
mass of  $D_s(2317)$ we
 find $\bar{E}_{B_s(1,\frac{1}{2},0)}=5634$ MeV. With this energy level
 we can then compute the levels of other P-wave states following 
 the same procedure 
 used for charmed mesons.
 The result is summarized in the first row for $\bar{E}$ in 
 Table \ref{b-mesons}.
An interesting feature of our estimation is that for $j=1/2$ the $B_s$ mesons have
almost equal or slightly smaller masses than 
their non-strange counterparts.

The masses for the P-wave bottom mesons can be obtained in a slightly
different way using the measured P-wave charmed meson masses. Identifying
the modified energy levels $\bar{E}$ for the charmed states  with $m_{\text 
expt.}$
in Table \ref{c-mesons} we can use Eq. ({\ref{charmbottomrel}) to obtain the
modified levels for the bottom mesons. The result is given in the second row for
$\bar{E}$ in Table \ref{b-mesons}. Although this method does not employ the 
chiral loop corrections explicitly the numbers
agree  well with those from the first approach.
This is an encouraging evidence for the consistency of our picture of the heavy-light
meson as a potential model bound state  with chiral cloud.

\begin{table}[t]
\begin{tabular}{c|cccccccc}\hline\hline
               &$D(1,\frac{1}{2},0)$&$D(1,\frac{1}{2},1)$&$D(1,\frac{3}{2},1)$&$D(1,\frac{3}{2},2)$&
	       $D_s(1,\frac{1}{2},0)$&$D_s(1,\frac{1}{2},1)$&$D_s(1,\frac{3}{2},1)$&$D_s(1,\frac{3}{2},2)$\\ \hline
$m_{\rm expt.}$&2308&2427&2422&2459&2317&2460&2535&2573 \\
$\bar E $      &2308&2421&2425&2468&2317&2435&2527&2573 \\ 
$E$            &2377&2490&2417&2460&2487&2605&2535&2581\\
\hline\hline
\end{tabular}
\caption{Modified energy levels $\bar E$ for P-wave
 charmed mesons. $E$
quoted from Ref. \cite{pe}. (Units
are in MeV.)}
\label{c-mesons}
\end{table}
\begin{table}[t]
\begin{tabular}{c|cccccccc}\hline\hline
               &$B(1,\frac{1}{2},0)$&$B(1,\frac{1}{2},1)$&$B(1,\frac{3}{2},1)$&$B(1,\frac{3}{2},2)$&
	       $B_s(1,\frac{1}{2},0)$&$B_s(1,\frac{1}{2},1)$&$B_s(1,\frac{3}{2},1)$&$B_s(1,\frac{3}{2},2)$\\ \hline
\multirow{2}{*}{$\bar E $}      &5637&5673&5709&5723&5634&5672&5798&5813 \\ 
               &5637&5679&5705&5713&5634&5697&5805&5812 \\ \hline
$E$            &5706&5742&5700&5714&5804&5842&5805&5820\\
\hline\hline
\end{tabular}
\caption{Modified energy levels $\bar E$ for P-wave
 bottom mesons. $E$
quoted from Ref. \cite{pe}. (Units are in MeV.)}
\label{b-mesons}
\end{table}

Now for D-wave  mesons we can use 
Eq. (\ref{diffrel2}) to compute the mass 
differences within the families. In Tables \ref{d-state1},\ref{d-state2}
the mass differences
$\Delta\equiv {\bar E}_{H(2,j,J)}-{\bar E}_{H(2,\frac{3}{2},1)}$ are
summarized. As in P-wave 
states there are no longer spin-orbit inversions
when the loop corrections are incorporated in.

\begin{table}[t]
\begin{tabular}{c|cccccc}\hline\hline
          &$D(2,\frac{3}{2},2)$&$D(2,\frac{5}{2},2)$&$D(2,\frac{5}{2},3)$&

$D_s(2,\frac{3}{2},2)$&$D_s(2,\frac{5}{2},2)$&$D_s(2,\frac{5}{2},3)$ \\ \hline
$\Delta$    &38&53&77&40&78&103\\
\hline\hline
\end{tabular}
\caption{Level differences in D-wave 
charmed mesons. (Units are in MeV.)}
\label{d-state1}
\end{table}

\begin{table}[t]
\begin{tabular}{c|cccccc}\hline\hline
          &$B(2,\frac{3}{2},2)$&$B(2,\frac{5}{2},2)$&$B(2,\frac{5}{2},3)$&

$B_s(2,\frac{3}{2},2)$&$B_s(2,\frac{5}{2},2)$&$B_s(2,\frac{5}{2},3)$ \\ \hline
$\Delta$    &12&33&41&13&59&67\\
\hline\hline
\end{tabular}
\caption{ Level differences in D-wave 
 bottom mesons. (Units are in MeV.)}
\label{d-state2}
\end{table}

\section{Conclusion}
We have calculated in relativistic potential model
the one loop chiral corrections for the energy
levels of the heavy-light mesons in P and D-wave
states with $j=3/2, 5/2$, 
and shown that the loop corrections can explain the absence of 
the spin-orbit inversions in  charmed mesons.
The disappearance of spin-orbit inversions by the
loop corrections is not confined to P-wave 
states only, as we have explicitly
shown with D-wave 
states, and appears to be a generic feature of the
potential model.
We proposed that the discrepancy between the
observed masses and potential model predictions for heavy-light mesons
can be remedied once  the chiral loop corrections are included, and this
allowed us to predict the masses for the P-wave 
bottom mesons which may be
tested in the near future.

\begin{acknowledgments}
We are very grateful to B.Y. Park for useful discussions.
This work was supported in part by 
Korea Research Foundation 
Grant (KRF-2005-015-C00107,KRF-2006-015-C00251).
\end{acknowledgments}



\end{document}